\documentclass[12pt]{article}

\usepackage{graphicx}
\usepackage{times}
\usepackage{color}
\definecolor{red}{rgb}{1,0,0}

\title{The world-wide air transportation network: Anomalous
centrality, community structure, and cities' global roles}

\author{R. Guimer\`a$^1$, S. Mossa$^2$, A. Turtschi$^1$, and
L. A. N. Amaral$^1$\\
\normalsize{$^1$Department of Chemical and Biological Engineering,}\\
\normalsize{Northwestern University, Evanston, IL 60208, USA}\\
\normalsize{$^2$European Synchrotron Radiation Facility,}\\
\normalsize{B.P. 220 F-38043 Grenoble, Cedex France}\\
}

\date{\today}

\begin{document}

\baselineskip 18pt

\maketitle


\begin{abstract}
We analyze the global structure of the world-wide air transportation
network, a critical infrastructure with an enormous impact on local,
national, and international economies. We find that the world-wide air
transportation network is a scale-free small-world network. In
contrast to the prediction of scale-free network models, however, we
find that the most connected cities are not necessarily the most
central, resulting in anomalous values of the centrality. We
demonstrate that these anomalies arise because of the multi-community
structure of the network. We identify the communities in the air
transportation network and show that the community structure cannot be
explained solely based on geographical constraints, and that
geo-political considerations have to be taken into account. We
identify each city's global role based on its pattern of inter- and
intra-community connections, which enables us to obtain scale-specific
representations of the network.
\end{abstract}
%
%



Like other critical infrastructures, the air transportation network
has enormous impact on local, national, and international
economies. It is thus natural that airports and national airline
companies are often times associated with the image a country or
region wants to
project~\cite{bissessur98,dana99,turton96,raguraman98}.  

The air transportation system is also responsible, indirectly, for the
propagation of diseases such as influenza and, recently, SARS. The air
transportation network thus plays for certain diseases a role that is
analogous to that of the web of human sexual contacts
\cite{liljeros01} for the propagation of AIDS and other
sexually-transmitted infections \cite{liljeros03,pastor01b}.

The world-wide air transportation network is responsible for the
mobility of millions of people every day. Almost 700 million
passengers fly each year, maintaining the air transportation system
ever so close to the brink of failure. For example, US and foreign
airlines schedule about 2,700 daily flights in and out of O'Hare
alone, more than 10\% of the total commercial flights in the
continental US, and more than the airport could handle even during a
perfect ``blue-sky'' day. Low clouds, for example, can lower landing
rates at O'Hare from 100 an hour to just 72 an hour, resulting in
delays and flight cancellations across the country. The failures and
inefficiencies of the air transportation system have large economic
costs; flight delays cost European countries 150 to 200 billion Euro
in 1999 alone \cite{delay00}.

These facts prompt several questions: What has led the system to this
point? Why can't we design a better system?  In order to answer these
questions, it is crucial to characterize the structure of the
world-wide air transportation network and the mechanisms responsible
for its evolution. The solution to this problem is, however, far from
simple. The structure of the air transportation network is mostly
determined by the concurrent actions of airline companies---both
private and national---that try, in principle, to maximize their
immediate profit.  However, the structure of the network is also the
outcome of numerous historical ``accidents'' arising from
geographical, political, and economic factors.

Much research has been done on the definition of models and algorithms
that enable one to solve problems of optimal network design
\cite{magnanti84,minoux89}. However, a world-wide, ``system'' level,
analysis of the structure of the air transportation network is still
lacking. This is an important caveat. Just as one cannot fully
understand the complex dynamics of ecosystems by looking at simple
food chains \cite{camacho02} or the complex behavior in cells by
studying isolated biochemical pathways \cite{kitano02,jeong00}, one
cannot fully understand the dynamics of the air transportation system
without a ``holistic'' perspective. Modern ``network analysis''
\cite{strogatz01,albert02,dorogovtsev02,newman03b,amaral04} provides
an ideal framework within which to pursue such a study.

We analyze here the world-wide air transportation network. We build a
network of 3883 locales, villages, towns, and cities with at least one
airport and establish links between pairs of locales that are
connected by non-stop passenger flights. We find that the world-wide
air transportation network is a small-world network~\cite{amaral00}
for which (i) the number of non-stop connections from a given city,
and (ii) the number of shortest paths going through a given city have
distributions that are scale-free. In contrast to the prediction of
scale-free network models, we find that the most connected cities are
not necessarily the most ``central,'' that is, the cities through
which most shortest paths go. We show that this surprising result can
be explained by the existence of several distinct ``communities''
within the air transportation network. We identify these communities
using algorithms recently developed for the study of complex networks,
and show that the structure of the communities cannot be explained
solely based on geographical constraints, and that geo-political
considerations must also be taken into account. The existence of
communities leads us to the definition of each city's global role,
based on its pattern of inter- and intra-community connections.

\section*{The data}

Many measures---including total number of passengers, total number of
flights, or total amount of cargo---quantifying the importance of the
world airports are compiled and publicized~\cite{ACI99}. We study here
the {\it OAG MAX\/} database\footnote{For more information on the OAG
MAX database, visit http://oagdata.com/solutions/max.aspx.}, which
comprises flight schedule data of more than 800 of the world's
airlines for the period November 1, 2000 to October 31, 2001. This
database is compiled by OAG, a division of Reed Business Information
Group, and includes all scheduled flights and scheduled charter
flights, both for big aircrafts---air carriers---and small
aircrafts---air taxis.

We focus our analysis on a network of cities, not of airports---for
example, the Newark, JFK and La Guardia airports are all assigned to
New York City.  We further restrict our analysis to {\it passenger
flights\/} operating in the time period November 1, 2000 to November
7, 2000.  Even though this data is more than four years old, the
resulting world-wide airport network is virtually indistinguishable
from the network one would obtain if using data collected today. The
reason is that air traffic patterns are strongly correlated with: (i)
socio-economic factors, such as population density and economic
development; and (ii) geo-political factors such as the distribution
of the continents over the surface of the Earth and the locations of
borders between states \cite{guimera04}. Clearly, the time scales
associated to changes in these factors are much longer than the lag in
the data we analyze here.

During the period considered, there are 531,574 unique non-stop
passenger flights, or flight segments, operating between 3883 distinct
cities. We identify 27,051 distinct city pairs having non-stop
connections. The fact that the database is highly redundant---that is,
that most connections between pairs of cities are represented by more
than one flight---adds reliability to our analysis. Specifically, the
fact that unscheduled flights are not considered does not mean, in
general, that the corresponding link between a certain pair of cities
is missing in the network, since analogous scheduled flights may still
operate between them. Similarly, even if some airlines have canceled
their flights between a pair of cities since November 2000, it is
highly unlikely that all of them have.

We create the corresponding adjacency matrix for this network, which
turns out to be almost symmetrical. The very minor asymmetry stems
from the fact that a small number of flights follow a ``circular''
pattern, i.e, a flight might go from A to B to C, and back to A. To
simplify the analysis, we symmetrize the adjacency matrix.

Further, we build regional networks for different geographic regions
(Table~\ref{t-regions}).  Specifically, we generate twenty-one
regional networks at different aggregation levels. At the
highest-aggregation level, we generate six networks; one each for
Africa, Asia and Middle East, Europe, Latin America, North America,
and Oceania.  For each of these regions, except for North America and
Oceania, we generate between two and five sub-networks. For instance,
the Asia and Middle East network is further subdivided into South
Asia, Central Asia, Southeast Asia, Northeast Asia, and Middle East.


\section*{Large-scale structure of the air transportation network}

A ubiquitous characteristic of complex networks is the so-called
``small-world'' property \cite{watts98}. In a small-world network,
pairs of nodes are connected by short paths as one expects for a
random graph \cite{bollobas01}. Crucially, nodes in small-world
networks also have a high-degree of cliquishness, as one finds in
low-dimensional lattices but not in random graphs.

In the air transportation network, the average shortest path length
$d$ is the average minimum number of flights that one needs to take to
get from any city to any other city in the world.  We find that for
the $719$ cities in the Asia and Middle East network $d=3.5$, and that
the average shortest path length between the $3663$ cities in the
giant component of the world-wide network is only about one step
greater, $d=4.4$. Actually, most pairs of cities (56\%) are connected
by four steps or less. More generally, we find that $d$ grows
logarithmically with the number $S$ of cities in the network, $d\sim
\log\,S$. This behavior is consistent with both random graphs and
small-world networks, but not with low-dimensional networks, for which
$d$ grows more rapidly with $S$.

Still, some pairs of cities are considerably further away from each
other than the average. The farthest cities in the network are Mount
Pleasant, in the Falkland Islands, and Wasu, in Papua New Guinea: To
get from one city to the other, one needs to take fifteen different
flights. From Mount Pleasant, one can fly to Punta Arenas, in Chile,
and from there to some hubs in Latin America. At the other end of the
path, from Wasu one needs to fly to Port Moresby, which requires a
unique sequence of eight flights. In the center of the path, between
Punta Arenas and Port Moresby, six different flights are needed. In
contrast with what happens the ends of the path, in the central region
of the path there are hundreds of different flight combinations, all
of them connecting Punta Arenas and Port Moresby in six steps.

The clustering coefficient $C$, which quantifies the local
cliquishness of a network, is defined as the probability that two
cities that are directly connected to a third city are also directly
connected to each other.  We find that $C$ is typically larger for the
air transportation network than for a random graph and that it
increases with size.  These results are consistent with the
expectations for a small-world network but not those for a random
graph. For the world-wide network, we find $C=0.62$ while its
randomization yields $C=0.049$. Therefore, we conclude that the air
transportation network is, as expected, a small-world network
\cite{amaral00}.

Another fundamental aspect in which real-world networks often deviate
from the random graphs typically considered in mathematical analysis
\cite{bollobas01} is the degree distribution, that is, the
distribution of the number of links of the nodes
\cite{barabasi99,amaral00,albert02}. In binomial random graphs, all
nodes have similar degrees, while many real world networks have some
nodes that are significantly more connected than others. Specifically,
many complex networks, termed scale-free, have degree distributions
that decay as a power law. A plausible mechanism for such a phenomenon
is preferential attachment \cite{barabasi99,amaral00}, that is, the
tendency to connect preferentially to nodes with already high degrees.

\begin{figure}[h]
\centerline{\includegraphics*[width=0.4\textwidth]{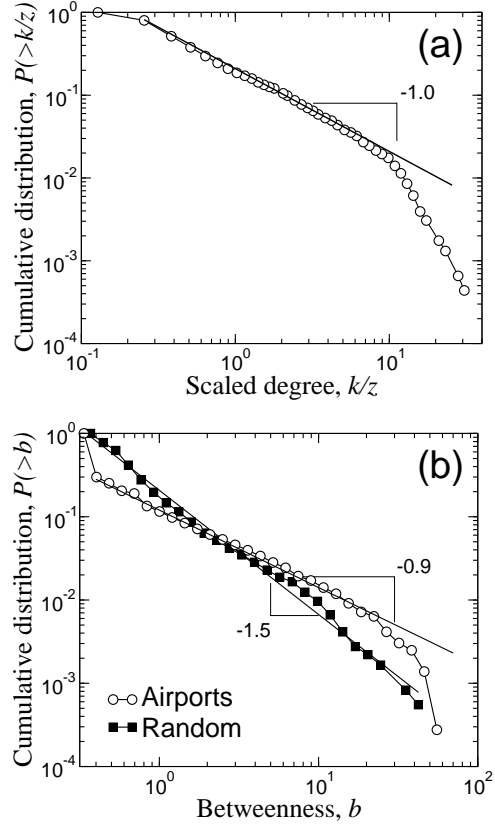}}
\caption{
Degree and betweenness distributions of the world-wide air transportation
network.
(a) Cumulative degree distribution plotted in double logarithmic
scale. The degree $k$ is scaled by the average degree $z$ of the
network. The distribution displays a truncated power law behavior with
exponent $\alpha = 1.0 \pm 0.1$.
(b) Cumulative distribution of normalized betweennesses plotted in
double logarithmic scale. The distribution displays a truncated power
law behavior with exponent $\nu = 0.9 \pm 0.1$. For a randomized
network with exactly the same degree distribution as the original air
transportation network, the betweenness distribution decays with an
exponent $\nu = 1.5 \pm 0.1$.  A comparison of the two cases clearly
shows the existence of an excessive number of large betweenness values
in the air transportation network.
}
\label{fig1}
\end{figure}

To gain greater insight into the structure and evolution of the air
transportation network, we calculate the degree distribution of the
cities. The degree of a city is the number of other cities to which it
is connected by a non-stop flight. In Fig.~\ref{fig1}a, we show the
cumulative degree distribution \footnote{The cumulative degree
distribution $P(>k)$ gives the probability that a city has $k$ or more
connections to other cities, and is defined as $P(>k) =
\sum_{k'=k}^\infty p(k')$, where $p(k)$ is the probability density
function. }
for the world-wide air transportation network. The data suggest that
$P(>k)$ has a truncated power-law decaying tail
\cite{barabasi99,amaral00}
\begin{equation}
P(>k) \propto k^{-\alpha}~f(k/k_{\times})
\label{e.truncated}
\end{equation}
where $\alpha = 1.0 \pm 0.1$ is the power law exponent, $f(u)$ is a
truncation function, and $k_{\times}$ is a crossover value that
depends on the size of the network.  The measured value of the
exponent $\alpha$ would imply that, as one increases the size of the
network, the average degree of the cities are also expected to
increase \cite{albert02}.


The degree of a node is a source of information on its
importance. However, the degree does not provide complete information
on the {\it role\/} of the node in the network. To start to address
this issue, we consider the ``betweenness centrality'' of the cities
comprising the world-wide air transportation network. The betweenness
$B_i$ of city $i$ is defined as the number of shortest paths
connecting any two cities that involve a transfer at city
$i$~\cite{freeman77,Newman01a,Newman01}. We define the normalized
betweenness as $b_i=B_i / \langle B \rangle$, where $\langle B
\rangle$ represents the average betweenness for the network. We plot,
in Fig.~\ref{fig1}b, the cumulative distribution $P(>b)$ of the
normalized betweenness for the world-wide air transportation
network. Our results suggest that the distribution of betweennesses
for the air transportation network obeys the functional form
\begin{equation}
P(>b) \propto b^{-\nu}~g(b/b_{\times})
\label{e.truncated2}
\end{equation}
where $\nu = 0.9 \pm 0.1$ is the power law exponent, $g(u)$ is a
truncation function, and $b_{\times}$ is a crossover value that
depends on the size of the network.


A question prompted by the previous results regarding the degree and
the centrality of cities is: ``Are the {\it most connected\/} cities
also the {\it most central\/}?'' To answer this question, we analyze
first the network obtained by randomizing the world-wide air
transportation network (Fig.~\ref{fig1}b).  We find that the
distribution of betweennesses still decays as a power law but, in this
case, with a much larger exponent value $\nu = 1.5 \pm 0.1$.  This
finding indicates the existence of anomalously large betweenness
centralities in the air transportation network.

\begin{figure}[h] 
\centerline{\includegraphics*[width=.4\textwidth]{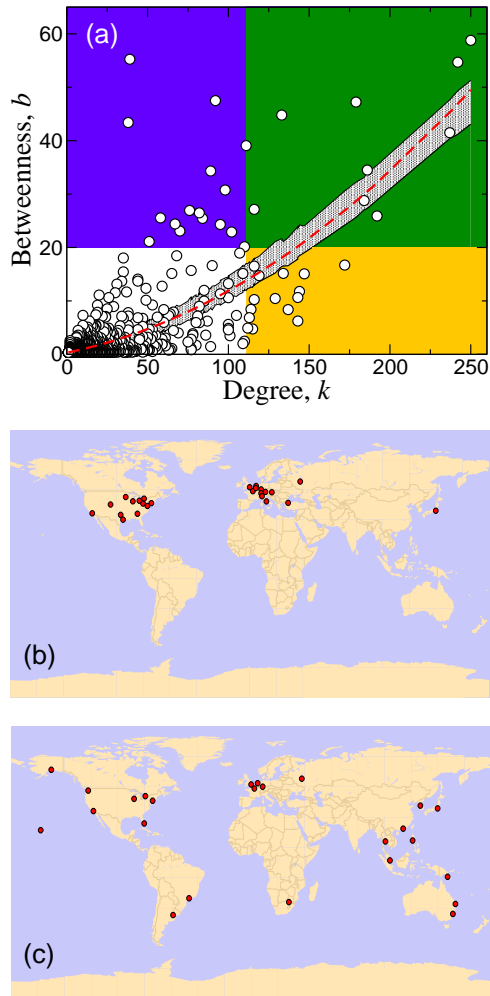}}
\caption
{
Most connected versus most central cities in the world-wide air
transportation network.
(a) Betweenness as a function of the degree for the cities in the
world-wide air transportation network (circles). For the randomized
network, the betweenness is well described as a quadratic function of
the degree (dashed line) with 95\% of all data falling inside the gray
region. In contrast to the strong correlation between degree and
betweenness found for randomized networks, the air transportation
network comprises many cities that are highly connected but have small
betweenness and, conversely, many cities with small degree and large
betweenness. We define a blue region containing the 25 most central
cities in the world, and a yellow region containing the 25 most
connected cities.  Surprisingly, we find there are only a few cities
with large betweenness and degree---green region (which is the
intersection of the blue and yellow regions).
(b) The 25 most connected cities in the world.
(c) The 25 most central cities in the world.
%
%
}
\label{fig3}
\end{figure}

For the randomized network, the degree of a node and its betweenness
centrality are strongly correlated---i.e., highly connected nodes are
also the most central (Fig.~\ref{fig3}a). In contrast, for the
world-wide air transportation network it turns out that there are
cities that are not hubs---i.e., have small degrees---but that
nonetheless have very large betweennesses (Fig. \ref{fig3}a).

To better illustrate this finding, we plot the 25 most connected
cities and contrast such a plot with another of the 25 most
``central'' cities according to their betweenness (Figs.~\ref{fig3}b
and c). While the most connected cities are located mostly in Western
Europe and North America, the most central cities are distributed
uniformly across all the continents. Significantly, each continent has
at least one central city, which is typically highly-connected when
compared to other cities in the continent---Johannesburg in Africa or
Buenos Aires and S\~{a}o Paulo in South America. Interestingly,
besides these cities with relatively large degree, there are
others---such as Anchorage (Alaska, U.S.) and Port Moresby (Papua New
Guinea)---that, despite having small degrees are among the most
central in the network (Table~1).

\section*{Degree-betweenness anomalies and multi-community networks}

Nodes with small degree and large centrality can be regarded as
anomalies. Other complex networks that have been described in the
literature, like the Internet \cite{vazquez02}, do not display such a
behavior, and nodes with the highest degree are also those with the
highest betweenness \cite{goh03}. It is, in principle, easy to
construct a network in which a node has small degree and large
centrality---think, for example, of a network formed by two
communities that are connected to one another through a single node
with only two links. The relevant question is, however, ``what general
and plausible mechanism would give rise to scale-free networks with
the obtained anomalous distribution of betweenness centralities?''

To answer this question it is useful to consider a region such as
Alaska.  Alaska is a sparsely populated, isolated region with a
disproportionately large---for its population size---number of
airports. Most Alaskan airports have connections only to other Alaskan
airports. This fact makes sense geographically.  However,
distance-wise it would also make sense for some Alaskan airports to be
connected to airports in Canada's Northern Territories.  These
connections are, however, absent.  Instead, a few Alaskan airports,
singularly Anchorage, are connected to the continental US.  The reason
is clear, the Alaskan population needs to be connected to the
political centers, which are located in the continental US, while
there are political constraints making it difficult to have
connections to cities in Canada, even to ones that are close
geographically \cite{guimera04}. It is now obvious why Anchorage's
centrality is so large.  Indeed, the existence of nodes with anomalous
centrality is related to the existence of regions with a high density
of airports but few connections to the outside. The degree-betweenness
anomaly is therefore ultimately related to the existence of
``communities'' in the network.

The unexpected finding of central nodes with low degree is a very
important one because central nodes play a key role in phenomena such
as diffusion and congestion \cite{guimera02b}, and in the cohesiveness
of complex networks \cite{holme02}. Therefore, our finding of
anomalous centralities points to the need to (i) identify the
communities in the air transportation network and (ii) establish new
ways to characterize the role of each city based on its pattern of
intra- and inter-community connections and not merely on its degree.

\begin{figure}[h]
\centerline{\includegraphics*[width=.9\textwidth]{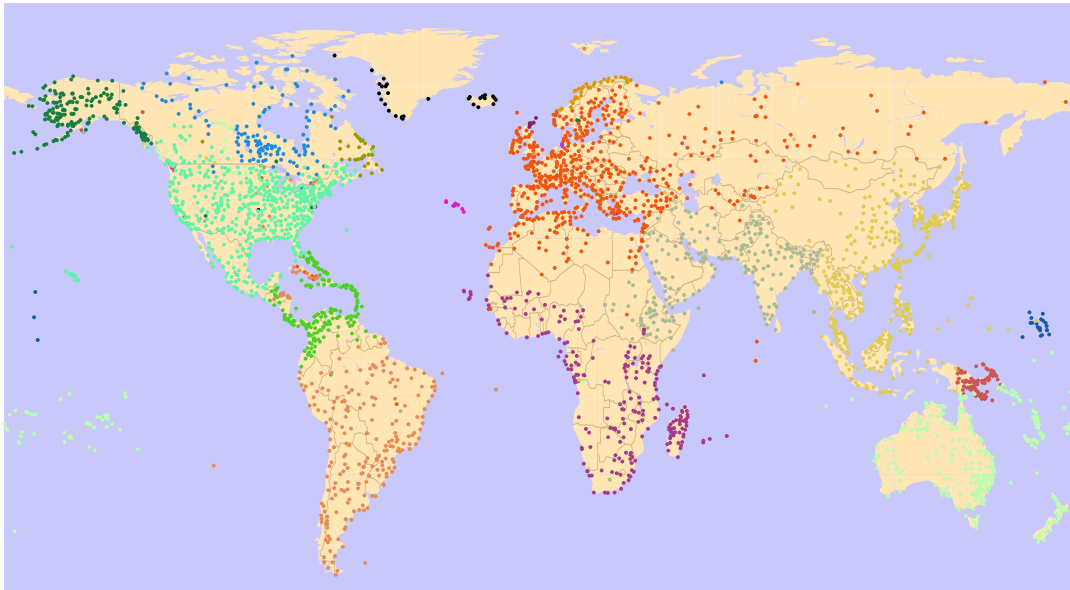}}
\caption{
  Communities in the giant component of the world-wide air
  transportation network. Each node represents a city and each color
  corresponds to a community.
}
\label{modules}
\end{figure}

\noindent
{\it Community structure---} To identify communities in the air
transportation network, we use the definition of modularity introduced
in Refs.~\cite{newman03,newman04}. The modularity of a given partition
of the nodes into groups is maximum when nodes that are densely
connected among them are grouped together and separated from the other
nodes in the network. To find the partition that maximizes the
modularity, we use simulated annealing
\cite{kirkpatrick83,guimera04c,guimera05,guimera05b}. We display in
Fig.~\ref{modules} the communities identified by our algorithm in the
world-wide air transportation network.\footnote{We do not know the
geographical coordinates of approximately 10\% of the 3663 cities in
the giant component of the world-wide air transportation
network. Those cities are not plotted in the map. Also, some small
cities may be misplaced due to duplications in the three-letter code
of the corresponding airport.}

As we surmised, both Alaska and Papua New Guinea form separate
communities.\footnote{Alaska and Papua New Guinea are small
communities compared to most of the others, which confirms the idea
that these are very well defined communities---otherwise, they would
be incorporated in a larger community. This fact is particularly
important taking into consideration that the community identification
algorithm does not take into account the betweenness of the nodes at
all.}
This fact explains the large betweenness centrality of Anchorage and
Port Moresby, as they provide the main links to the outside world for
the other cities in their communities.

Another significant result is that even though geographical distance
plays a clear role in the definition of the communities, the
composition of some of the communities cannot be explained by purely
geographical considerations. For example, the community that contains
most cities in Europe also contains most airports in Asian
Russia. Similarly, Chinese and Japanese cities are mostly grouped with
cities in the other countries in Southeast Asia, but India is mostly
grouped with the Arabic Peninsula countries and with countries in
Northeastern Africa.  These facts are consistent with the important
role of political factors in determining community structure
\cite{guimera04}.

\noindent
{\it Global role of cities---} We characterize the role of each city
in the air transportation network based on its pattern of intra- and
inter-community connections. We first distinguish nodes that play the
role of hubs in their communities from those that are non-hubs. Note
that cities like Anchorage are hubs in their communities but they are
not hubs if one considers all the nodes in the network. Thus, we
define the within-community degree of a node.  If $\kappa_{i}$ is the
number of links of node $i$ to other nodes in its community $s_i$,
$\overline{\kappa}_{s_i}$ is the average of $\kappa$ over all the
nodes in $s_i$, and $\sigma_{\kappa_{s_i}}$ is the standard deviation
of $\kappa$ in $s_i$, then
\begin{equation}
z_i = \frac{\kappa_i - \overline{\kappa}_{s_i}}{\sigma_{\kappa_{s_i}}}
\end{equation}
is the so-called $z$-score. The within-community degree $z$-score
measures how ``well-connected'' node $i$ is to other nodes in the
community.

We then distinguish nodes based on their connections to nodes in
communities other than their own. For example, two nodes with the same
$z$-score will play different roles if one of them is connected to
several nodes in other communities while the other is not. We define
the participation coefficient $P_i$ of node $i$ as
\begin{equation}
P_i=1-\sum_{s=1}^{N_M}\left(\frac{\kappa_{is}}{k_i} \right)^2
\end{equation}
where $\kappa_{is}$ is the number of links of node $i$ to nodes in
community $s$, and $k_i$ is the total degree of node $i$. The
participation coefficient of a node is therefore close to one if its
links are uniformly distributed among all the communities, and zero if
all its links are within its own community.

We hypothesize that the role of a node can be determined, to a great
extent, by its within-module degree and its participation coefficient
\cite{guimera05,guimera05b}. We define heuristically seven different
``universal roles,'' each one corresponding to a different region in
the $zP$ phase-space. According to the within-module degree, we
classify nodes with $z \ge 2.5$ as module hubs and nodes $z<2.5$ as
non-hubs. Both hub and non-hub nodes are then more finely
characterized by using the values of the participation coefficient
\cite{guimera05,guimera05b}.

We divide non-hub nodes into four different roles: (R1) {\it
ultra-peripheral nodes}, i.e., nodes with all their links within their
module ($P \le 0.05$); (R2) {\it peripheral nodes}, i.e., nodes with
most links within their module ($0.05<P \le 0.62$); (R3) {\it non-hub
connector nodes}, i.e., nodes with many links to other modules
($0.62<P \le 0.80$); and (R4) {\it non-hub kinless nodes}, i.e., nodes
with links homogeneously distributed among all modules ($P>0.80$).

We divide hub nodes into three different roles: (R5) {\it provincial
hubs}. i.e., hub nodes with the vast majority of links within their
module ($P \le 0.30$); (R6) {\it connector hubs}, i.e., hubs with many
links to most of the other modules ($0.30<P \le 0.75$); and (R7) {\it
kinless hubs}, i.e., hubs with links homogeneously distributed among
all modules ($P>0.75$).

\begin{figure}[t!]
\centerline{\includegraphics*[width=.9\textwidth]{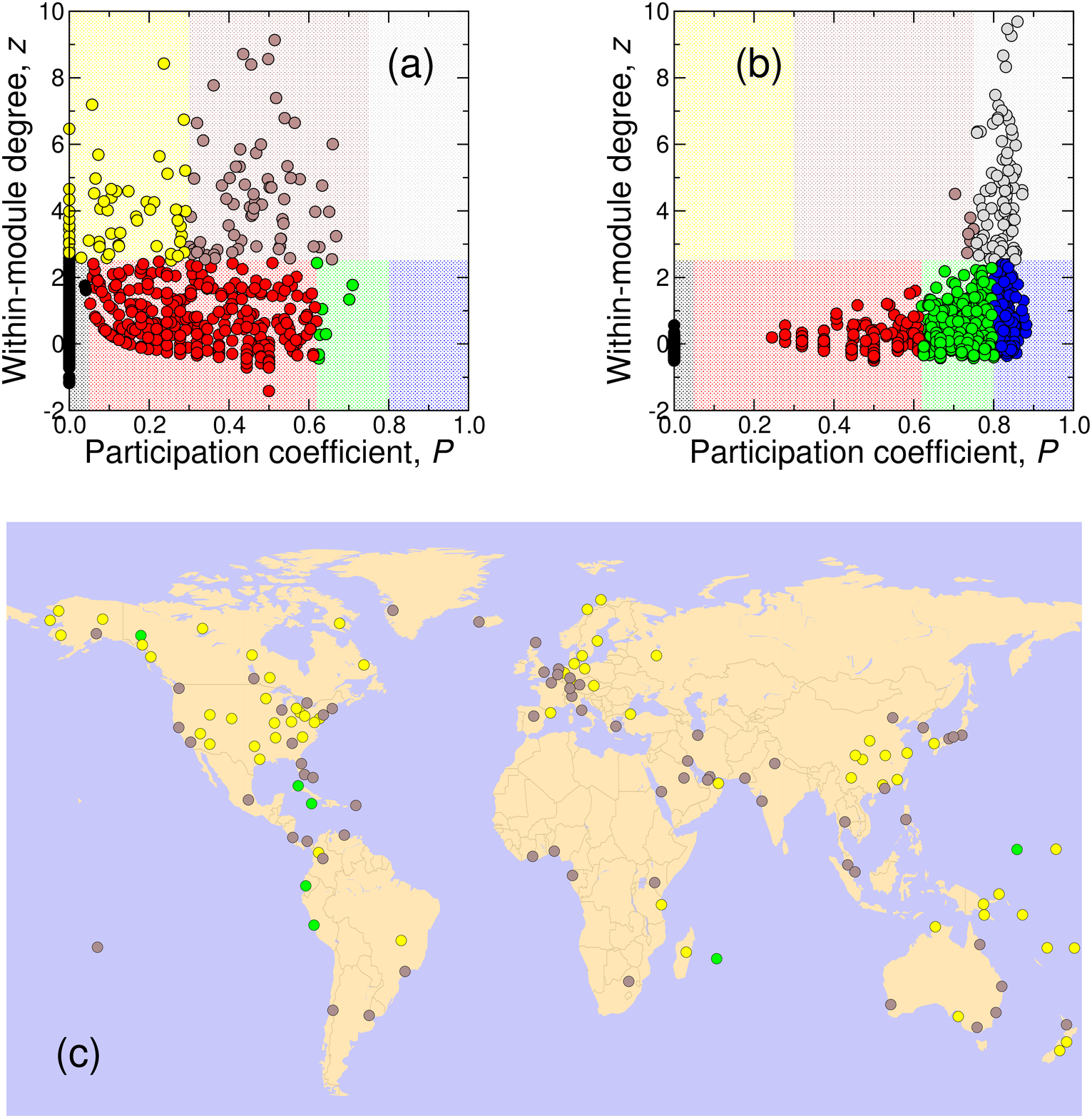}}
\caption{
  Toward a scale-specific representation of the world-wide air
  transportation network.
  (a) Each point in the $zP$ phase-space corresponds to a city, and
  different colors indicate different roles. Most cities are
  classified as ultra-peripheral (black) or peripheral (red) nodes. A
  small number of non-hub nodes play the role of connectors
  (green). We find approximately equal fractions of provincial
  (yellow) and connector (brown) hubs.
  (b) Same as (a) but for a randomization of the air transportation
  network. The absence of communities manifests itself in that most
  hubs become kinless hubs (gray) and in the appearance of kinless
  non-hubs (blue).
  (c) Non-hub connectors (green), provincial hubs (yellow), and
  connector hubs (brown) in the world-wide air transportation
  network.
}
\label{f-roles}
\end{figure}

For each city in the world-wide air transportation network, we
calculate its within-community degree $z_i$ and its participation
coefficient $P_i$. Then, we assign each city a role according to the
definitions above (Figs.~\ref{f-roles}a and c).  Significantly, 95.4\%
of the cities in the world-wide air transportation network are
classified as either peripheral or ultra-peripheral. Additionally,
there is a small fraction of non-hub connectors (0.5\%). This result
suggests that cities which are not hubs in their respective
communities rarely have links to many other communities in the air
transportation network. This situation is in stark contrast to what
happens in some biological networks, in which non-hub connectors seem
to be relatively frequent and to play an important role
\cite{guimera05}.

The remaining 4.1\% of the nodes are hubs.  We find approximately
equal fractions of provincial and connector hubs. The former include
cities that, for historical, political, or geographical reasons, are
comparatively not well-connected to other communities. Examples are
Denver, Philadelphia, and Detroit, in North America; Stuttgart,
Copenhagen, Istanbul, and Barcelona, in the community formed by
Europe, North Africa and the former Soviet Union; Adelaide and
Christchurch in Oceania; Brasilia in South America; Fairbanks and
Juneau in Alaska; and the already discussed case of Port
Moresby. Connector hubs include the most recognizable airport hubs in
the word: Chicago, New York, Los Angeles, and Mexico City in North
America; Frankfurt, London, Paris, and Rome in Europe; Beijing, Tokyo,
and Seoul in the South-Eastern Asian community; Delhi, Abu Dhabi, and
Kuwait in the community comprising India, the Arabic Peninsula and
North-Eastern Africa; Buenos Aires, Santiago, and S\~ao Paulo in South
America; Melbourne, Auckland, and Sydney in Oceania; and Anchorage in
Alaska.

The fractions of cities with each role in the world-wide air
transportation network contrast with the corresponding fractions in a
randomization of the network (Fig.~\ref{f-roles}b). In this case, the
community identification algorithm still yields certain communities,
but the network lacks ``real'' community structure. The identification
of roles enables one to realize that these communities are somehow
artificial. Indeed, many cities are either kinless hubs or kinless
non-hubs due to the absence of a real community structure, and the
network contains essentially no provincial or connector hubs.

\section*{Discussion}

We carried out a ``systems'' analysis of the structure of the
world-wide air transportation network. The study enables us to unveil
a number of significant results. The world-wide air transportation
network is a small-word network in which: (i) the number of non-stop
connections from a given city, and (ii) the number of shortest paths
going through a given city have distributions that are
scale-free. Surprisingly, the nodes with more connections are not
always the most central in the network. We hypothesize that the origin
of such a behavior is the multi-community structure of the network. We
find the communities in the network and demonstrate that their
structure can only be understood in terms of both geographical and
political considerations.

Our analysis of the community structure of the air transportation
network is important for two additional reasons. First, it allow us to
identify the most efficient ways in which to engineer the structure of
the network.  Specifically, having identified the communities, one can
identify which ones are poorly connected and the ways to minimize that
problem.  Second, cities that connect different communities play a
disproportionate role in important dynamic processes such as the
propagation of infections such as SARS. As, we show, finding the
communities is the first step toward identifying these cities.

The existence of communities and the understanding that different
cities may have very different impacts on the global behavior of the
air transportation system, call for the definition of the role of each
city. We address this issue by classifying cities into seven roles,
according to their patterns of inter- and intra-community
connections. We find that most of the nodes (95\%) are peripheral,
that is, the vast majority of their connections are within their own
communities. We also find that nodes that connect different
communities are typically hubs within their own community, although
not necessarily global hubs. This finding is in stark contrast with
the behavior observed in certain biological networks, in which non-hub
connectors are more frequent \cite{guimera05}.

The fact that different networks seem to be formed by nodes with
network-specific roles points to the more general question of what
evolutionary constraints and pressures determine the topology of
complex networks, and how the presence or absence of specific roles
affects the performance of these networks.

%
\bibliographystyle{pnas}


\begin{center}
{\bf Acknowledgments}
\end{center}

We thank A. Arenas, A. Barrat, M. Barth\'el\'emy,
A. D\'{\i}az-Guilera, A.~A. Moreira, R. Pastor-Satorras,
M. Sales-Pardo, D. Stouffer, and A. Vespignani for stimulating
discussions and helpful suggestions. We also thank OAG for making
their electronic database of airline flights available to us, and
Landings.com for providing us with the geographical coordinates of the
world airports.

\clearpage

\centerline{\Large \bf Tables}

%
\begin{table}[h]
\begin{center}
\begin{tabular}{lr}
\hline
Region & Number of cities\\
\hline
Africa & 364 \\
Asia and Middle East & 719 \\
Europe & 691 \\
Latin America & 523 \\
North America & 1064 \\
Oceania & 522 \\
\hline
\end{tabular}
\end{center}
\caption{Number of cities with airports by major geographic region.}
\label{t-regions}
\end{table}

%
\begin{table}[h]
\begin{center}
\begin{tabular}{llcrr}
\hline
Rank & City & $b$ & $b/b_{ran}$ & Degree\\
\hline
1 & Paris & 58.8 & 1.2 & 250\\
\color{red}{2} & \color{red}{Anchorage} & \color{red}{55.2} & \color{red}{16.7} & \color{red}{39}\\
3 & London & 54.7 & 1.2 & 242\\
\color{red}{4} & \color{red}{Singapore} & \color{red}{47.5} & \color{red}{4.3} & \color{red}{92}\\
5 & New York & 47.2 & 1.6 & 179\\
\hline
6 & Los Angeles & 44.8 & 2.3 & 133\\
\color{red}{7} & \color{red}{Port Moresby} & \color{red}{43.4} & \color{red}{13.6} & \color{red}{38}\\
8 & Frankfurt & 41.5 & 0.9 & 237\\
9 & Tokyo & 39.1 & 2.7 & 111\\
10 & Moscow & 34.5 & 1.1 & 186\\
\hline
\color{red}{11} & \color{red}{Seattle} & \color{red}{34.3} & \color{red}{3.3} & \color{red}{89}\\
\color{red}{12} & \color{red}{Hong Kong} & \color{red}{30.8} & \color{red}{2.6} & \color{red}{98}\\
13 & Chicago & 28.8 & 1.0 & 184\\
14 & Toronto & 27.1 & 1.8 & 116\\
\color{red}{15} & \color{red}{Buenos Aires} & \color{red}{26.9} & \color{red}{3.2} & \color{red}{76}\\
\hline
\color{red}{16} & \color{red}{S\~ao Paulo} & \color{red}{26.5} & \color{red}{2.8} & \color{red}{82}\\
17 & Amsterdam & 25.9 & 0.8 & 192\\
\color{red}{18} & \color{red}{Melbourne} & \color{red}{25.5} & \color{red}{4.5} & \color{red}{58}\\
\color{red}{19} & \color{red}{Johannesburg} & \color{red}{25.4} & \color{red}{2.6} & \color{red}{84}\\
\color{red}{20} & \color{red}{Manila} & \color{red}{24.4} & \color{red}{3.5} & \color{red}{67}\\
\hline
\color{red}{21} & \color{red}{Seoul} & \color{red}{24.3} & \color{red}{2.1} & \color{red}{95}\\
\color{red}{22} & \color{red}{Sydney} & \color{red}{23.1} & \color{red}{3.2} & \color{red}{70}\\
\color{red}{23} & \color{red}{Bangkok} & \color{red}{22.9} & \color{red}{1.8} & \color{red}{102}\\
\color{red}{24} & \color{red}{Honolulu} & \color{red}{21.1} & \color{red}{4.4} & \color{red}{51}\\
\color{red}{25} & \color{red}{Miami} & \color{red}{20.1} & \color{red}{1.4} & \color{red}{110}\\
\hline
\end{tabular}

\end{center}
\caption{The 25 most central cities in the world-wide air
transportation network, ordered according to their normalized
betweenness. We also show the ratio of the actual betweenness of the
cities to the betweenness that they have after randomizing the
network. Cities displayed in red are not among the 25 most connected.}
\label{table1}
\end{table}
%

\clearpage

%

\centerline{\Large \bf Figures}

%
%

%
%
%
%

%
%

\end{document}